\documentclass[%
 reprint,
superscriptaddress,
 amsmath,amssymb,
 aps,
]{revtex4-2}

\usepackage{graphicx}
\graphicspath{ {./images/} }
\usepackage{floatrow}
\usepackage[dvipsnames]{xcolor}
\usepackage{dcolumn}%
\usepackage{bm}%
\usepackage[caption=false]{subfig}
\usepackage{hyperref}%

\begin{document}

\newcommand{\kBT}{k_\mathrm{B}T}
\newcommand{\tm}{\mathrm}
\newcommand{\Vc}{V_\mathrm{c}}
\newcommand{\Vb}{V_\mathrm{b}}
\newcommand{\nuc}{\nu_\mathrm{c}}
\newcommand{\nub}{\nu_\mathrm{b}}
\newcommand{\nus}{\nu_\mathrm{s}}
\newcommand{\mub}{\mu_\mathrm{b}}
\newcommand{\muc}{\mu_\mathrm{c}}
\newcommand{\vect}{\boldsymbol}
\newcommand{\diff}{\mathrm{d}}
\newcommand{\PhiTot}{\Phi_{\mathrm{tot}}}
\newcommand{\Domega}{\Delta\omega}
\newcommand{\DomegaEff}{\Domega_\mathrm{eff}}
\newcommand{\figref}[1]{Fig.~\ref{#1}}
\newcommand{\appref}[1]{Appendix~\ref{#1}}
\newcommand{\Eqref}[1]{Eq.~\eqref{#1}}
\newcommand{\Eqsref}[1]{Eqs.~\eqref{#1}}
\newcommand{\Kd}{K_\mathrm{d}}
\newcommand{\cd}{c_\mathrm{d}}
\newcommand{\DZ}[1]{\textcolor{purple}{DZ: #1}}
\newcommand{\RR}[1]{\textcolor{teal}{RR: #1}}
\newcommand{\GW}[1]{\textcolor{olive}{GW: #1}}
\newcommand{\olig}{\ell}

\preprint{APS/123-QED}

\title{Binding and dimerization control phase separation in a compartment}%

\author{Riccardo Rossetto}
 \author{Gerrit Wellecke}
\affiliation{Max Planck Institute for Dynamics and Self-Organization, Am Faßberg 17, 37077 Göttingen, Germany}%
\affiliation{University of Göttingen, Institute for the Dynamics of Complex Systems, Friedrich-Hund-Platz 1, 37077 Göttingen, Germany}%
\author{David Zwicker}%
 \email{david.zwicker@ds.mpg.de}
\affiliation{Max Planck Institute for Dynamics and Self-Organization, Am Faßberg 17, 37077 Göttingen, Germany}%
\date{\today}%

\begin{abstract}
    Biological cells exhibit a hierarchical spatial organization, where various compartments harbor condensates that form by phase separation. Cells can control the emergence of these condensates by affecting compartment size, the amount of the involved molecules, and their physical interactions. While physical interactions directly affect compartment binding and phase separation, they can also cause oligomerization, which has been suggested as a control mechanism. Analyzing an equilibrium model, we illustrate that dimerization amplifies compartment binding and phase separation, which reinforce each other. This nonlinear interplay can also induce multistability, which provides additional potential for control. Our work forms the basis for deriving thermodynamically consistent kinetic models to understand how biological cells can regulate phase separation in their compartments.

\end{abstract}

\maketitle

\section{Introduction}

Biological cells face the difficult challenge of controlling the spatial organization of their biomolecules.
They partly meet this challenge by compartmentalization, where lipid membranes or other well-defined structures physically separate the cytosol into chemically-distinct regions.
Within each compartment, spontaneous accumulation of biomolecules, in particular via phase separation~\cite{Hyman2014, Banani2017}, can impart further organization.
Examples for these hierarchical organizations encompass compartments of various spatial dimension; see \figref{fig:compartment_full_model_schematic}a.
One example for a linear, 1D compartment is the synaptonemal complex (SC), a liquid-crystalline compartment forming between homologous chromosomes during meiosis~\cite{Rog2017}, which binds molecules that can then diffuse along it and accumulate in small foci~\cite{Morgan2021,Zhang2021c,Durand2022}.
Protein accumulation on DNA might be another example for a similar 1D structure~\cite{DNA_ph_sep}.
Less speculative is the formation of protein-rich domains in 2D lipid membranes~\cite{Banjade2014}.
These domains are thought to form by phase separation~\cite{Case2019, Kamatar2024, Litschel2024, Wan2024}, and the involved proteins can exchange with the bulk cytosol~\cite{Weakly2024} or with separated membrane regions~\cite{Lipowsky2013,Giomi2019, Giomi2020}.
Finally, there are many examples of 3D compartments, either membrane-bound, like the nucleus, or membrane-less, like the nucleolus, which can harbor phase-separating proteins~\cite{nucleolus}.
These examples show that cells employ an hierarchical organization, combining spontaneous phase separation with binding to existing compartments.

Biological cells also face the difficulty to respond to external stimuli and deploy appropriate responses.
In our context, this implies that cells control phase separation in compartments.
While they could regulate the amount of material to control whether phase separation takes place~\cite{ Banani2017,Hyman2014}, a faster, and presumably more efficient, approach is to regulate the physical properties of the involved molecules, e.g., by post-translational modifications~\cite{Snead2019}.
On the one hand, molecular properties directly affect the affinity toward a compartment and phase separation.
On the other hand, specific physical interactions between proteins might lead to the formation of small oligomers, which could affect these processes.
Indeed, there is experimental evidence that oligomerization regulates protein domains on membranes~\cite{Bland2023, Illukkumbura2023, Tschirpke2023_Liedewij,Litschel2024}.
The aim of this paper is to disentangle the respective contributions of compartment binding, unspecific interactions, and oligomerization on phase separation in compartments.

Biological cells are alive, implying that many processes are driven away from equilibrium, which could also affect protein domains in compartments.
Previous work has described such systems using kinetic models~\cite {Lang2022, Turner2005}, reaction-diffusion equations~\cite{Burkart2022, Gross2019, Brauns2023_Liedewij, Brauns2021}, and coupled Cahn-Hilliard equations~\cite {Caballero2023,Foret2005}.
However, these phenomenological models do not take into account thermodynamic constraints, so they cannot reveal the precise influence of the physical properties of the involved molecules.
Thermodynamically consistent models have been derived for the specific case of wetting and pre-wetting on a membrane with surface binding~\cite{Zhao2021, Zhao2024}, but a more general theory is missing.
As a step toward such a general theory, we here focus on the behavior within the compartment by %
discussing an equilibrium theory that captures the interplay between compartment binding, unspecific interactions enabling phase separation, and dimerization as an example of oligomerization; see \figref{fig:compartment_full_model_schematic}b.

\section{Equilibrium model of phase separation in a compartment}
\begin{figure}[t]
  \centering
  {\includegraphics[width= \columnwidth]{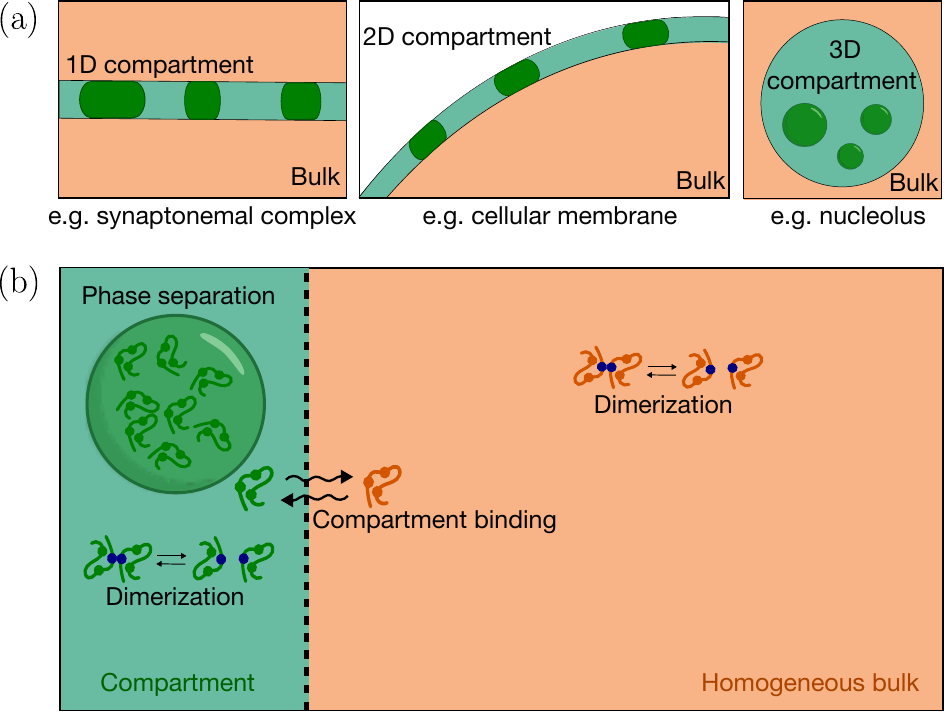}}\quad%
  \caption{\textbf{Schematic picture of the equilibrium model}. 
  (a) Examples of phase separation in compartments of different dimensions coupled to a large bulk.
  (b) Our model describes potential dimerization ($\omega_2$), compartment binding ($\Domega$), and unspecific interactions ($\chi$) leading to phase separation in the compartment, while the large bulk stays homogeneous.
}
\label{fig:compartment_full_model_schematic}
\end{figure}

Our model describes solute molecules in a compartment of fixed volume~$\Vc$ coupled to a larger bulk of volume $\Vb$.
To account for different morphologies (see \figref{fig:compartment_full_model_schematic}a), the dimension~$n$ of the compartment can differ from the dimension~$m$ of the bulk, implying that $\Vc$ and $\Vb$ may have different units.
Each subsystem is filled with a liquid mixture consisting of a solvent and a solute, which can form dimers.
The state of the compartment is characterized by the volume fraction fields $\phi_1(\vect r)$ and $\phi_2(\vect r)$ for the solute monomers and dimers, respectively, whereas the solvent fraction is $1-\phi_1-\phi_2$.
Analogously, the state of the bulk is described by fields $\psi_1(\vect r)$ and $\psi_2(\vect r)$.
Material conservation implies the constraint
\begin{equation}
\label{eq:conservation}
	\frac{1}{\nuc} \int_{\Vc} \bigl(\phi_1 + \phi_2\bigr) \diff^n r +
   \frac{1}{\nub}\int_{\Vb}  \bigl(\psi_1 + \psi_2\bigr)  \diff^m r   =N_{\tm{tot}}  \;, 
\end{equation}
where $N_\mathrm{tot}$ is the total number of solute molecules with molecular volumes $\nuc$ and $\nub$ in the compartment and bulk, respectively.
Introducing the fraction of solute in the compartment, $\phi = \phi_1 + \phi_2$, and the solute fraction in the bulk, $\psi = \psi_1 + \psi_2$, material conservation becomes
\begin{equation}
	\label{eq:conservation_homogeneous}
   \eta \bar\psi + \bar\phi = \PhiTot
   \;,
\end{equation}
where $\bar\phi= \Vc^{-1}\int\phi \, \diff^n r$ and $\bar\psi= \Vb^{-1}\int\psi \, \diff^m r$ are the average solute fractions in the respective subsystems.
The key parameter  $\eta = (\nu_\tm{c} V_\tm{b})/(\nu_\tm{b}V_\tm{c} )$ quantifies the size of the bulk relative to the compartment, where here size is measured by the maximal number of solutes that can fit in the respective subsystem.
Without loss of generality, we assume that the bulk is larger, $\eta > 1$, essentially defining which subsystem is the bulk.
Typical values are $\eta\sim115$ in the PAR system of \textit{C. elegans}~\cite{Goehring2011, Bland2023} and $\eta\sim90$ for the synaptonemal complex in \textit{A. thalania} (private communication), so we use $\eta=100$ in this paper.
The final parameter in \Eqref{eq:conservation_homogeneous} is the fixed normalized total solute amount  $\PhiTot= N_{\text{tot}}\nu_\tm{c}/\Vc$, which denotes the fraction of the compartment occupied by solute if the bulk was empty.
Note that this fraction can exceed $1$ if there is more material than can fit in the compartment.

The equilibrium behavior of the system is governed by the total free energy,
\begin{align}
	F = \int_{\Vc} f_\tm{c}(\phi_1,\phi_2)  \diff^n  r + \int_{\Vb} f_\tm{b}(\psi_1,\psi_2)  \diff^m  r
	\label{eqn:free_energy}
	\;,
\end{align}
where we neglect coupling terms describing wetting  since we focus on the behavior inside the compartment.
Such couplings could also affect phase separation when the compartment's dimension is lower than the bulk's, but this typically just leads to rescaled parameters; see \Eqref{eqn:coupling_rescaling} in Appendix~\ref{sec:boundary_coupling}.
We thus focus on the bulk effects inside the subsystems, which are only coupled via material exchange.
The equilibrium of this situation is governed by the respective free energy densities accounting for dimerization and phase separation,
\begin{subequations}
    \label{eq:free_energy_full}
    \begin{align}
     f_\tm{c}  &=   \frac{k_\tm{B}T} { \nu_\tm{c}}\Bigl[
     	\phi_{1} \ln(\phi_{1}) +\frac{\phi_{2}}{2} \ln(\phi_{2}) +  (1-\phi) \ln(1-\phi_{\tm{}}) 
\notag\\ &\quad +\omega_{1,\tm{c}}\phi_{1} + \omega_{2,\tm{c}} \phi_{2} +\chi\phi_{\tm{}}(1-\phi_{\tm{}})\Bigr] 
	\;, \qquad \text{and}
    \label{eq:free_energy_full_mem}
\\[8pt]
        f_\tm{b}  &= \frac{k_\tm{B}T} { \nu_\tm{b}}\Bigl[
        \psi_{1} \ln(\psi_{1}) +\frac{\psi_{2}}{2} \ln(\psi_{2}) +   (1-\psi_{\tm{}}) \ln(1-\psi_{\tm{}})
\notag\\ &\quad +\omega_{1,\tm{b}}\psi_{1} + \omega_{2,\tm{b}} \psi_{2} +\chi \psi_{\tm{}}(1-\psi) \Bigr]
    \label{eq:free_energy_full_bulk}
    	\;.
    \end{align}
\end{subequations}
For each expression, the respective first three terms describe translational entropies, whereas the other terms account for enthalpic contributions of monomers (fourth term) and dimers (fifth term), as well as an unspecific interaction between all solute molecules captured by the Flory parameter~$\chi$~\cite{flory, Huggins1941}. 
For simplicity, we assume that solute monomers and solvent molecules have the same molecular size, e.g., because they are both macromolecules.
To clearly separate the influence of physical interactions and limit the number of physical parameters, we consider the case where unspecific interactions are the same in both subsystems, and that dimerization and unspecific interactions are independent.
Likewise, we consider the case where dimerization and compartment binding are decoupled, so that the total energy decreases by~$\omega_2$ when a solute molecule becomes part of a dimer, independent of the subsystem it is in.
This also implies that the energy $\Domega$  gained when a solute molecule enters the compartment is independent of whether this molecule is part of a dimer.
These choices imply $\omega_{1,\tm{b}}=0$, $\omega_{2,\tm{b}}=-\omega_2$, $\omega_{1,\tm{c}}=-\Domega$, and $\omega_{2,\tm{c}}=-\omega_2 - \Domega$.

\floatsetup[figure]{style=plain,subcapbesideposition=top}

The behavior of the system is governed by the balance of chemical potentials derived from \Eqsref{eq:free_energy_full} together with the constraint given by \Eqref{eq:conservation_homogeneous}.
This implies that neither the dimensions of the subsystems nor their absolute volumes affect the behavior, which is instead governed by the following five parameters:
$\eta$ determines the relative sizes of the subsystems, $\PhiTot$ sets the total amount of material, and the three interaction energies $\omega_2$, $\Domega$, and $\chi$ govern the tendency of the solute to dimerize, be located in the compartment, and phase separate, respectively; see \figref{fig:compartment_full_model_schematic}b.

\section{Results}

\subsection{Dimerization regulates compartment binding}
\begin{figure}[t]
 \centering
 {\includegraphics[width=\columnwidth]{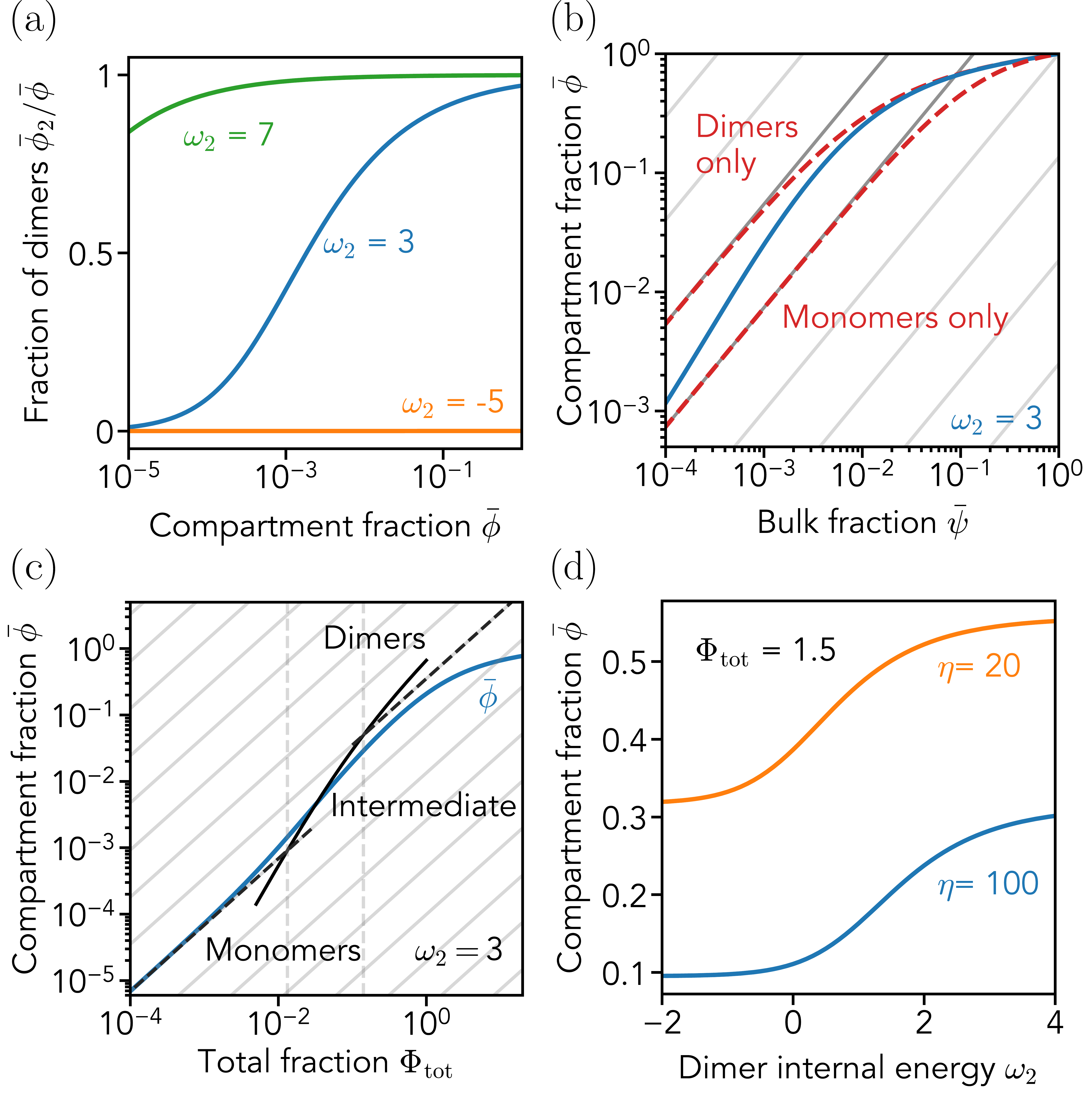} }
   
\caption{\textbf{Dimerization regulates compartment binding}.
(a) Fraction~$\bar\phi_2$ of dimers normalized to solute compartment fraction~$\bar\phi$ as as a function of $\bar\phi$ for a single compartment for various dimerization energies $\omega_2$.
(b) $\bar\phi$ in the compartment as a function of the solute fraction~$\bar\psi$ in the bulk for $\omega_2=3$.
The predictions for systems exclusively comprising monomers and dimers are shown as red dashed lines.
(c) $\bar\phi$ as a function of the total fraction~$\PhiTot$ in the system for $\omega_2=3$. The three scaling regimes discussed in the main text are highlighted by black lines.
(d) $\bar\phi$ as a function of $\omega_2$ for various $\eta$ at $\PhiTot=1.5$.
(a)--(d) Additional model parameters are $\eta=100$ and $\Domega=2$.
}
\label{fig:dimer_binding}
 \end{figure}

We first study the interplay between dimerization and compartment binding, neglecting unspecific interactions ($\chi=0$).
In this case, phase separation cannot take place and both subsystems remain homogeneous, so the free energy~$F$ given by \Eqref{eqn:free_energy} only depends on the average fractions $\bar\phi_1$, $\bar\phi_2$, $\bar\psi_1$, and $\bar\psi_2$.
To minimize $F$ under the material-conservation constraint given by \Eqref{eq:conservation_homogeneous}, we first minimize the two terms in \Eqref{eqn:free_energy} separately, which corresponds to equilibrating the dimerization reaction.
We can then express the monomer fraction $\bar\phi_1$ and the dimer fraction $\bar\phi_2$ in the compartment as a function of the solute compartment fraction $\bar\phi$,
\begin{align}
	\label{eqn:dimerization_equilibrium}
	\bar\phi_1 & = \frac{(1+4 K_\tm{d} \bar\phi)^{\frac12}-1}{2 K_\tm{d}} \;,
&
	\bar\phi_2 &= \bar\phi - \bar\phi_1
	\;,
\end{align}
introducing the dimerization constant $\Kd=e^{2\omega_2+1}$ governing the dimerization equilibrium, $\bar\phi_2/\bar\phi_1^2 = \Kd$; see Appendix~\ref {app:dim_eq}.
\Eqref{eqn:dimerization_equilibrium} implies that the compartment is dominated by monomers in the limit $\Kd\bar\phi\ll1$, whereas it contains mostly dimers when $\Kd\bar\phi\gg1$.
Moreover, $\bar\phi_2$ increases faster than $\bar\phi_1$ as a function of the solute compartment fraction~$\bar\phi$, implying that the ratio of dimers increases with $\bar\phi$; see \figref{fig:dimer_binding}a.
Analogous expressions apply for the fractions $\bar\psi_1$, $\bar\psi_2$, and $\bar\psi$ in the bulk.
For both subsystems, we can then express the free energy density as a function of $\bar\phi$ and $\bar\psi$ alone.
Since these two quantities are linked by the constraint given by \Eqref{eq:conservation_homogeneous}, we can express $F$ as a function of only the solute bulk fraction $\bar\psi$.
Minimizing $F$ with respect to $\bar\psi$ then yields the global equilibrium, where the dimerization in each subsystem and compartment binding are equilibrated.

We start by discussing the limiting cases of mostly monomers ($\omega_2\rightarrow -\infty$).
In this case, we have $\bar\phi_1 = \bar\phi$ and $\bar\psi_1 = \bar\psi$, whereas $\bar\phi_2=\bar\psi_2=0$. %
If the bulk is dilute ($\bar\psi_1\ll1$), we find $\bar\phi_1/ \bar\psi_1\approx(1-\bar\phi_1) e^{\Domega }$.
If the compartment is also dilute ($\bar\phi_1 \ll 1$), this implies that the ratio $\bar\phi_1/ \bar\psi_1$ is given by $e^{\Domega }$; see \figref{fig:dimer_binding}b.
However, if the overall amount of material increases sufficiently ($\PhiTot > \frac12 + \eta e^{-\Domega}$), the compartment is more than half filled ($\bar\phi_1>\frac12$), and the ratio  $\bar\phi_1/ \bar\psi_1$ significantly decreases below $e^{\Domega}$.
For much larger $\PhiTot$, the compartment saturates, so that excess amount is constrained to the bulk.
In the opposing limit of mostly dimers ($\omega_2\rightarrow\infty$), we find $\bar\phi_1=\bar\psi_1=0$, and $\bar\phi_2/\bar\psi_2 = (1-\bar\phi_2)^2 e^{2\Domega}$,  implying an increased affinity toward the compartment compared to the monomeric case ($\bar\phi_2/\bar\psi_2 > \bar\phi_1/\bar\psi_1$); see \figref{fig:dimer_binding}b.
Note that this increased affinity can be interpreted in two complementary pictures: %
Focusing on individual particles, dimerization reduces their translational entropy, thus requiring less binding energy to concentrate them in the compartment.
In contrast, if we focus on the number concentrations of monomers and dimers, we simply find that the binding energy of dimers is twice that of monomers, explaining their increased affinity.
Taken together, binding and dimerization increase the fractions $\phi_i$ in the compartment until saturation is reached.

We next focus on the intermediate case, where monomers and dimers can both be present in significant amounts.
For instance, in the case of $\omega_2=3$, \figref{fig:dimer_binding}b shows that the solute fractions $\bar\phi$ and $\bar\psi$ in the two subsystems transition from the monomeric limit to the dimeric limit when the total fraction $\PhiTot$ is increased.
This transition is a consequence of the preferred binding toward the compartment and the non-linear response of dimerization:
If the total fraction~$\PhiTot$ is low, both subsystems contain mostly monomers; see \Eqref{eqn:dimerization_equilibrium} and \figref{fig:dimer_binding}a.
Since the fraction in the compartment is larger than the fraction in the bulk ($\bar\phi>\bar\psi$), the compartment contains more dimers than the bulk, and will thus eventually transition into a state dominated by dimers while the bulk is still dominated by monomers.
Finally, for sufficiently large $\PhiTot$, the bulk will also contain mostly dimers and the transition is completed.
These three phases are summarized in \figref{fig:dimer_binding}c, indicating that we can predict the transition points by appropriate equilibration of the limiting cases; see \appref{app:transition_regimes}.
Note that the saturation of the compartment implies another non-linearity for very large $\PhiTot$.
Taken together, the interplay between dimerization and compartment binding leads to non-linear behavior, which has been observed experimentally~\cite{Bland2023}.
This behavior suggests that dimerization, as well as the relative size~$\eta$, can be used to control compartment binding. %

\floatsetup[figure]{style=plain,subcapbesideposition=top}

\subsection{Dimerization facilitates phase separation}

\begin{figure}[t]
  \centering
  {\includegraphics[width= \columnwidth]{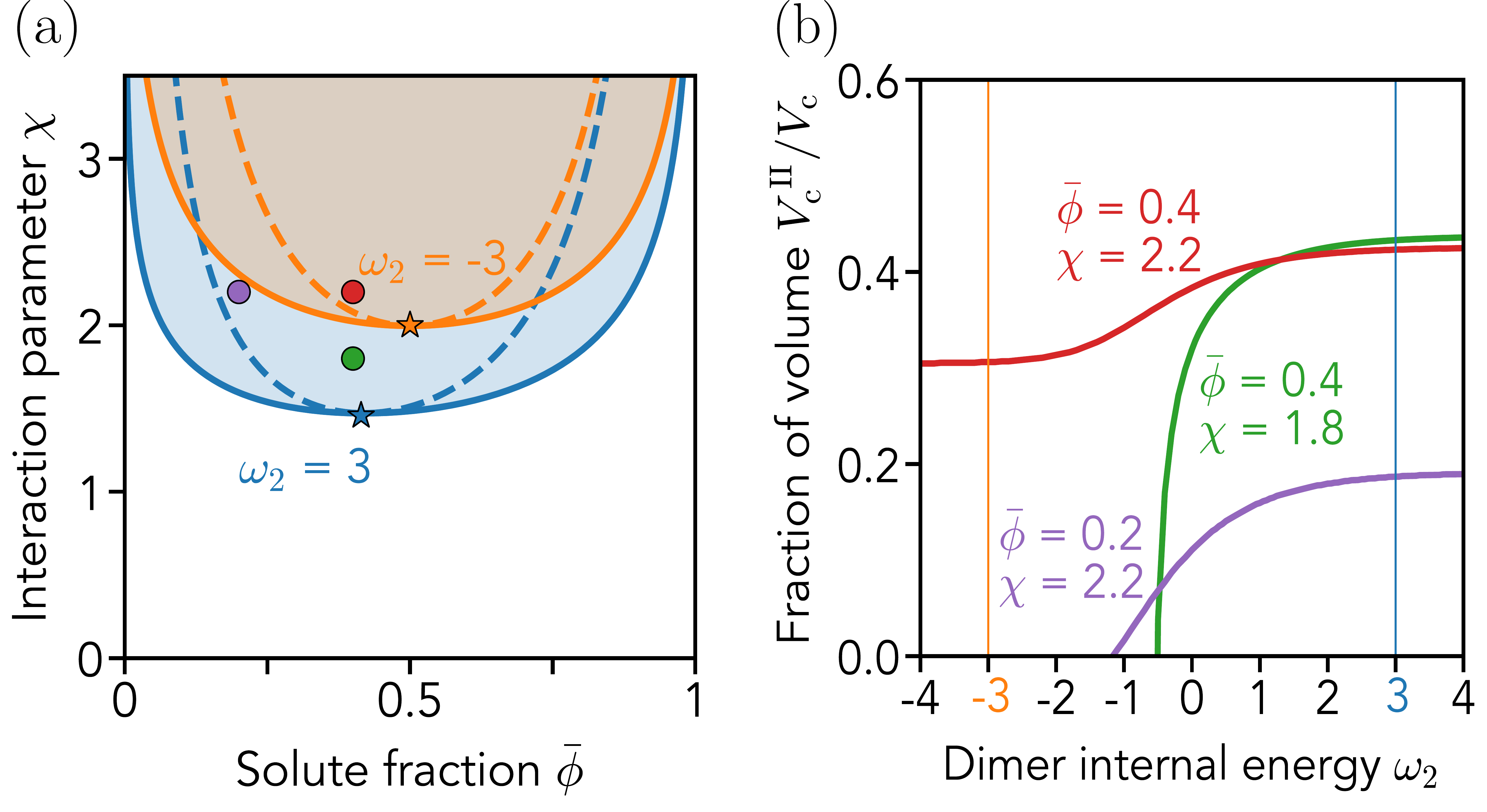}} 

  \caption{\textbf{Dimerization facilitates phase separation}.
      (a)~Phase diagrams as a function of solute fraction~$\bar\phi$  and interaction strength~$\chi$ for the two indicated dimerization energies $\omega_2$.
  Phase-separated states are stable above the solid binodal lines, whereas the homogeneous state is stable below the dashed spinodal lines.
  The critical points are marked by stars.
  (b) Fraction of volume~$V^\mathrm{II}_\mathrm{c}$ of the dense phase relative to the compartment volume $\Vc$ as a function of $\omega_2$ for various $\chi$ and $\bar\phi$ indicated in the plot and marked in panel~a.
  Vertical lines correspond to the values of $\omega_2$ shown in panel~a.
	}
  \label{fig:dimers+ph_sep}
\end{figure}

We next investigate how dimerization affects phase separation in the simple situation where the bulk is negligible ($\Domega \rightarrow\infty$), so we only focus on the compartment.
Since we now allow for phase separation ($\chi\neq0$), the compartment can be heterogenous and its state is described by the fields $\phi_1(\vect r)$ and $\phi_2(\vect r)$.
However, the dimerization equilibrium, $\delta F/\delta \phi_1 = \delta F/\delta \phi_2$, holds everywhere, allowing us to express these fields in terms of solute fraction $\phi(\vect r)$, analogously to \Eqref{eqn:dimerization_equilibrium}; see Appendix~\ref {app:dim_eq}.
Consequently, the total free energy is given by an integral over a free energy density that only depends on $\phi(\vect r)$, which is analogous to standard binary phase separation~\cite{Weber_2019,Zwicker2022}.
In particular, the compartment can separate into a dense and a dilute phase with identical dimerization equilibrium, $\phi_2/\phi_1^2 = \Kd$, if the interaction parameter $\chi$ is sufficiently large.
\figref{fig:dimers+ph_sep}a shows that larger $\chi$ is required to induce phase separation for smaller $\omega_2$ (smaller $\Kd$).
This effect can be understood by analyzing the homogeneous state $\phi(\vect r) = \bar\phi$, which becomes unstable in the spinodal region, $\chi > \chi_\mathrm{sp}$, with
\begin{equation}
    \chi_{\tm{sp}}=\frac{1}{4\bar\phi} \bigg(\frac{1}{ (1+4 K_\tm{d} \bar\phi)^{\frac12}}+\frac{1+\bar\phi}{1-\bar\phi} \bigg)
    \label{eqn:spinodal}
    \;;
\end{equation}
see Appendix~\ref {app:dim_eq}.
This expression shows that $\chi_\mathrm{sp}$ increases for smaller $\Kd=e^{2\omega_2+1}$.
This effect is stronger for smaller $\bar\phi$, implying that smaller $\omega_2$ move the spinodal region toward large solute fractions; see \figref{fig:dimers+ph_sep}a.
This behavior can also be understood from the limiting case of only monomers ($\omega_2\rightarrow-\infty$) and only dimers ($\omega_2 \rightarrow \infty$):
In terms of the Flory-Huggins theory~\cite{Huggins1941} the only difference between these two cases is the contribution to translational entropy, which is halved for the dimers.
Consequently, a smaller enthalpic gain by the unspecific interaction~$\chi$ can offset the entropic loss of concentrating particles in a phase.
The minimal value of $\chi$ necessary to obtain phase separation can be calculated by minimizing $\chi_\mathrm{sp}(\bar\phi)$ given by \Eqref{eqn:spinodal}, which corresponds to the critical point (stars in  \figref{fig:dimers+ph_sep}a).
For monomers ($\omega_2\rightarrow-\infty$), we obtain $\chi_\mathrm{sp}^\mathrm{min}=2$, whereas $\chi_\mathrm{sp}^\mathrm{min}=(3+2 \sqrt{2})/4 \simeq 1.46$ for dimers ($\omega_2 \rightarrow \infty$).
Taken together, this analysis shows that the dimerization energy $\omega_2$ can control phase separation and droplet size; see \figref{fig:dimers+ph_sep}b.

\floatsetup[figure]{style=plain,subcapbesideposition=top}

\subsection{Sequestration by the bulk suppresses phase separation in the compartment} \label{subsec:C}
We next study phase separation in the compartment coupled to the bulk without dimerization (only monomers, $\omega_2\rightarrow-\infty$).
We focus on the case where particles prefer the compartment ($\Domega>0$) since in the opposing case the smaller compartment would be negligible compared to the bulk.
In this case, the average fraction~$\bar\phi$ in the compartment is larger than the fraction~$\bar\psi$ in the bulk, favoring phase separation in the compartment.
The bulk would only phase separate if the total fraction~$\PhiTot$ is very large, at which point the compartment is saturated.
In this case, the compartment would only sequester a fixed amount of material, but not influence phase separation further.
More generally, at most one of two coupled subsystems can exhibit phase separation in equilibrium~\cite{Lipowsky2013, Giomi2019, Giomi2020, Zhao2024}.
This results from Gibbs phase rule~\cite{Gibbs1876}, which allows at most two phases per subsystem if they are not in contact; enabling material exchange (but keeping the volumes fixed)  adds one constraint, so the total phase count is three and not four.

Since only one subsystem can exhibit phase separation, we focus on the case where the compartment is favored ($\Domega>0$) and the bulk remains homogeneous with a fraction $\bar\psi$.
In this case, the bulk can be approximated as a dilute mixture, so the chemical potentials in the compartment, $\mu_\mathrm{c} =\nuc \partial f_\mathrm{c}/\partial \phi$, and the bulk, $\mu_\mathrm{b} =\nub  \partial f_\mathrm{b}/\partial \bar\psi$, read
\begin{subequations}
    \label{eq:free_energy_phase_comp}
    \begin{align}  \label{eq:chempot_comp_phcomp}
        \mu_\tm{c}(\phi) &= k_\mathrm{B}T \Bigl[ \ln\left(\frac{\phi}{1-\phi}\right) -\Domega+\chi (1 -2\phi) \Bigr]
\;,\\
    \label{eq:chempot_bulk_phcomp}
         \mu_\tm{b}(\bar\psi) &\approx k_\mathrm{B}T \Bigl[  \ln(\bar\psi) + \chi \Bigr]\;, 
    \end{align}
\end{subequations}
for $\bar\psi\ll1$. 
Since the assumption that the bulk is dilute could be violated in some cases, we will check it a posteriori.
Using material conservation given by \Eqref{eq:conservation_homogeneous} to eliminate $\bar\psi$ in \Eqref{eq:chempot_bulk_phcomp}, and shifting both chemical potentials by $\ln\eta - \chi$, we find
\begin{subequations}
    \label{eq:free_energy_phase_comp_eff}
    \begin{align}  \label{eq:chempot_comp_phcomp_eff}
       \mu^\mathrm{eff}_\tm{c}(\phi) &= k_\mathrm{B}T \Bigl[ \ln\left(\frac{\phi}{1-\phi}\right) -\DomegaEff -2\phi\chi \Bigr]
\;,\\
    \label{eq:chempot_bulk_phcomp_eff}
         \mu^\mathrm{eff}_\tm{b}(\bar\phi) &\approx k_\mathrm{B}T \Bigl[   \ln( \PhiTot -  \bar\phi ) \Bigr]
         \;,
    \end{align}
\end{subequations}
where $\DomegaEff = \Domega - \ln\eta$ captures the enthalpy and entropy gained when a solute particle moves from the bulk to the compartment. 
It shows that the relative size~$\eta$ of the subsystems plays a similar role to the compartment affinity~$\Domega$.
In particular, in the limit of a small bulk ($\eta\rightarrow0$), we find $\DomegaEff\rightarrow\infty$, so that all solute particles go to the compartment ($\bar\phi \rightarrow \PhiTot$).
In this case, the bulk becomes negligible, the total mass in the compartment is conserved, and the behavior in the compartment is governed by ordinary phase separation in the canonical ensemble.
In the contrasting case of a large bulk ($\eta\rightarrow\infty$), implying $\DomegaEff\rightarrow-\infty$,  we also need to increase the total fraction $\PhiTot$ to have solutes in the compartment.
This implies $\PhiTot \gg \bar\phi$, so that $\bar\phi$ is negligible in \Eqref{eq:chempot_bulk_phcomp_eff}, implying constant $\mu_\mathrm{b}$, which corresponds to a grand-canonical ensemble for the compartment.
The interesting case of a bulk of finite size thus interpolates between these two traditional ensembles.

\begin{figure}[t]
\centering
{\includegraphics[width=\columnwidth]{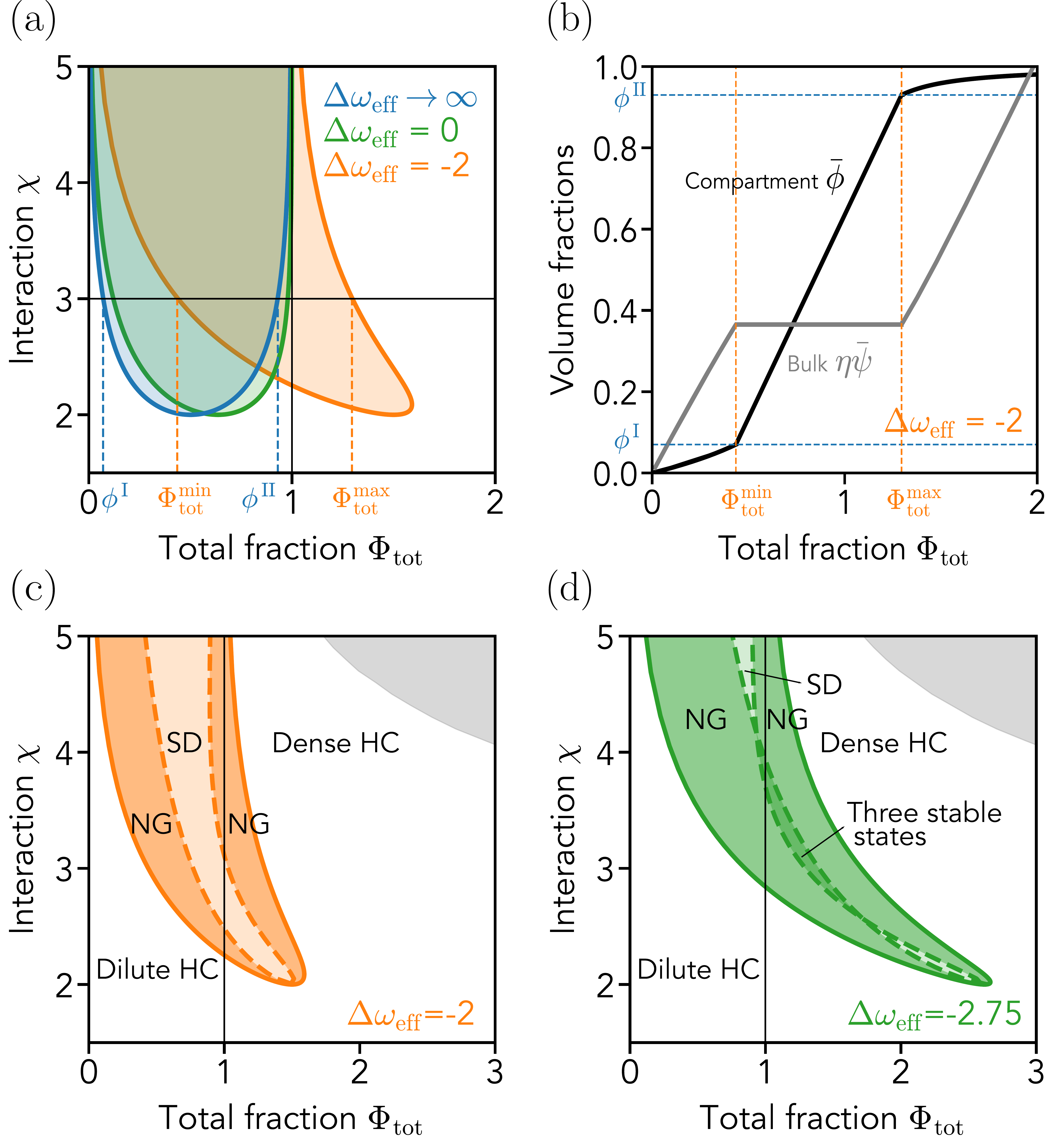}}
       
\caption{\textbf{Sequestration by the bulk suppresses phase separation in the compartment}.
(a) Phase diagrams as a function of the total fraction~$\PhiTot$ and interaction strength~$\chi$ for two  effective affinities $\DomegaEff=\Domega - \ln\eta$.
The compartment can phase separate within the colored regions.
The limit of strong binding (blue) recovers the canonical phase diagram, determining the coexisting volume fractions $\phi^\mathrm{I}$ and $\phi^\mathrm{II}$ (plot shows example for $\chi=3$).
Weaker binding (green and orange) moves the region to larger $\PhiTot$ and the intersections with the line $\chi=3$ now give the region $\PhiTot \in (\PhiTot^\mathrm{min}, \PhiTot^\mathrm{max})$ where phase separation takes place, whereas the coexisting volume fractions still correspond to $\phi^\mathrm{I/II}$.
(b) Average volume fractions $\bar\phi$ in the compartment and normalized fraction $\eta\bar\psi$ in the bulk as a function of $\PhiTot$ for $\DomegaEff=-2$ and $\chi=3$.
(c, d) Phase diagrams including spinodals (dashed lines) for two values of $\DomegaEff$, separating spinodal decomposition (SD) from the nucleation-and-growth (NG) region.
White regions denote a globally stable homogeneous configuration (HC), whereas HCs are only linearly stable in the darker region between the dashed and solid lines, leading to multistability in the central region of panel (d).
The bulk exhibits phase separation in the gray region for $\eta=100$. %
}
\label{fig:phase_diags_phsep}
\end{figure}

\figref{fig:phase_diags_phsep}a shows how the phase diagram depends on the effective affinity~$\DomegaEff$.
In the limit of large $\DomegaEff$ (blue lines), we indeed recover the canonical phase diagram of a binary system since the bulk is almost empty $\bar\psi\approx0$, whereas the solute fraction~$\bar\phi$ in the compartment approaches the total fraction $\PhiTot$.
Phase separation can thus take place in the binodal region, where $\PhiTot$ is between the fractions~$\phi^\mathrm{I}$ and $\phi^\mathrm{II}$ of the dilute and dense phase in the compartment, $\PhiTot \in (\phi^\mathrm{I}, \phi^\mathrm{II})$.
For smaller $\DomegaEff$ the entire phase diagram shifts toward larger $\PhiTot$ (orange lines), essentially because a significant fraction of solute molecules are in the bulk.
In particular, phase separation only occurs in the compartment when $\PhiTot \in (\PhiTot^\mathrm{min}, \PhiTot^\mathrm{max})$; see \figref{fig:phase_diags_phsep}a for an example at $\chi=3$.
However, the coexisting fractions in the dilute and dense phase in the compartment are still given by $\phi^\mathrm{I}$ and $\phi^\mathrm{II}$ (see \appref{app:ph_sep_comp}), and thus need to be read off the phase diagram corresponding to the canonical ensemble (blue lines).
In essence, the phase diagram as a function of $\PhiTot$ determines what total fractions lead to phase separation, whereas the coexisting volume fractions $\phi^\mathrm{I/II}$ need to be determined from the canonical phase diagram as a function of $\bar\phi$.
We show in \appref{app:ph_sep_comp} that the difference between these fractions are
\begin{align}
	\label{eqn:binodal_shift}
	\PhiTot^\mathrm{min} - \phi^\mathrm{I} 
	= \PhiTot^\mathrm{max} - \phi^\mathrm{II} 
	= e^{-\Domega_\mathrm{eff} -\chi}
	\;,
\end{align}
because the bulk sequesters a constant fraction of material when the compartment phase separates; see \figref{fig:phase_diags_phsep}b.
The fraction of material sequestered by the bulk thus decreases with larger effective affinity $\DomegaEff$ and larger interaction strength $\chi$, so that the phase diagram approaches the canonical limit for large $\DomegaEff+\chi$.
In essence, $\Domega$ and $\chi$ determine how much material ends up in the compartment, where phase coexistence then determines the fractions $\phi^\mathrm{I/II}$.
Changing $\PhiTot$ then only changes the relative sizes of the two phases according to the lever rule, leaving all fractions unchanged.
The fraction of volume of the dense phase  reads
\begin{equation}
  \frac{V_\tm{c}^{\tm{II}}}{V_\tm{c}}= \frac{1}{\phi^{\tm{II}}-\phi^{\tm{I}}}\Bigl( \Phi_{\text{tot}} - \phi^{\tm{I}} - e^{-\DomegaEff -\chi} \Bigr)
  \;,
\end{equation}
showing that the dense phase in the compartment increases when there is more material (larger $\PhiTot$) and larger effective affinity $\DomegaEff$.

Bulk sequestration changes the phase diagram qualitatively.
For instance, we find a re-entrant phase transition when changing the interaction parameter~$\chi$.
\figref{fig:phase_diags_phsep}a shows an example for $\DomegaEff=-2$ (orange lines) where phase separation is only possible in a narrow regime of $\chi$ when $\PhiTot$ is equal to the marked $\PhiTot^\mathrm{max}$.
This is essentially due to an increasing fraction~$\bar\phi$ in the compartment when $\chi$ increases, caused by the lower chemical potential; see \Eqref{eq:chempot_comp_phcomp_eff}.
Similarly, $\bar\phi$ increases monotonically (and $\bar\psi$ decreases accordingly) for larger affinity $\Domega$ or smaller size ratio~$\eta$.
In contrast, the total fraction $\PhiTot$ affects the system differently; see \figref{fig:phase_diags_phsep}b.
While $\bar\phi$ and $\bar\psi$ increase with $\PhiTot$ if the compartment is homogeneous (outside the binodal), only the size of the dense phase increases, but the fractions stay constant, when the compartment phase separates, which is akin to noise buffering described in simple phase separation~\cite{Klosin2020}.
Taken together, we show that the affinity $\Domega$, the size ratio~$\eta$, and the interaction strength~$\chi$ can be used to control phase separation in the compartment together with the total fraction~$\PhiTot$.

We next discuss the stability of homogeneous states to distinguish nucleation and growth regimes from spinodal decomposition.
\figref{fig:phase_diags_phsep}c shows that the spinodal (dashed orange line) also moves toward larger total fractions~$\PhiTot$ for smaller $\DomegaEff$, similarly to the binodals discussed above.
In this case, the entire phase diagram is thus qualitatively similar to the phase diagram in the canonical limit ($\DomegaEff\rightarrow\infty$).
In contrast, phase diagrams for smaller $\DomegaEff$, where the compartment is less favored, can be qualitatively different; see \figref{fig:phase_diags_phsep}d.
In this case, the spinodal decomposition region is reduced dramatically, and the two nucleation-and-growth regimes overlap.
In this overlap region (dark green), there are three stable states:  the phase separated state has the lowest free energy, but there are also two locally stable homogeneous states.
This is because the binding equilibrium, $\mu_\mathrm{c}(\bar\phi) = \mu_\mathrm{b}(\bar\phi)$ using \Eqsref{eq:free_energy_phase_comp_eff}, leads to a non-linear equation with three solutions, of which two are stable; see \appref{app:ph_sep_comp}.
Note that the bulk can also exhibit phase separation when the compartment is densely occupied (at large $\PhiTot$ and $\chi$; gray region in \figref{fig:phase_diags_phsep}), but we do not analyze this case further since it is roughly equivalent to phase separation in the bulk with part of the material sequestered by the compartment.
In the more interesting case where the compartment exhibits phase separation, we showed that a medium-sized bulk can affect phase separation qualitatively, leading to the effective binding affinity  $\DomegaEff = \Domega - \ln\eta$ as a central control parameter.

\floatsetup[figure]{style=plain,subcapbesideposition=top}

\subsection{Dimerization and bulk sequestration jointly regulate compartment phase separation}

\begin{figure}[t]
\centering
{\includegraphics[width= \columnwidth]{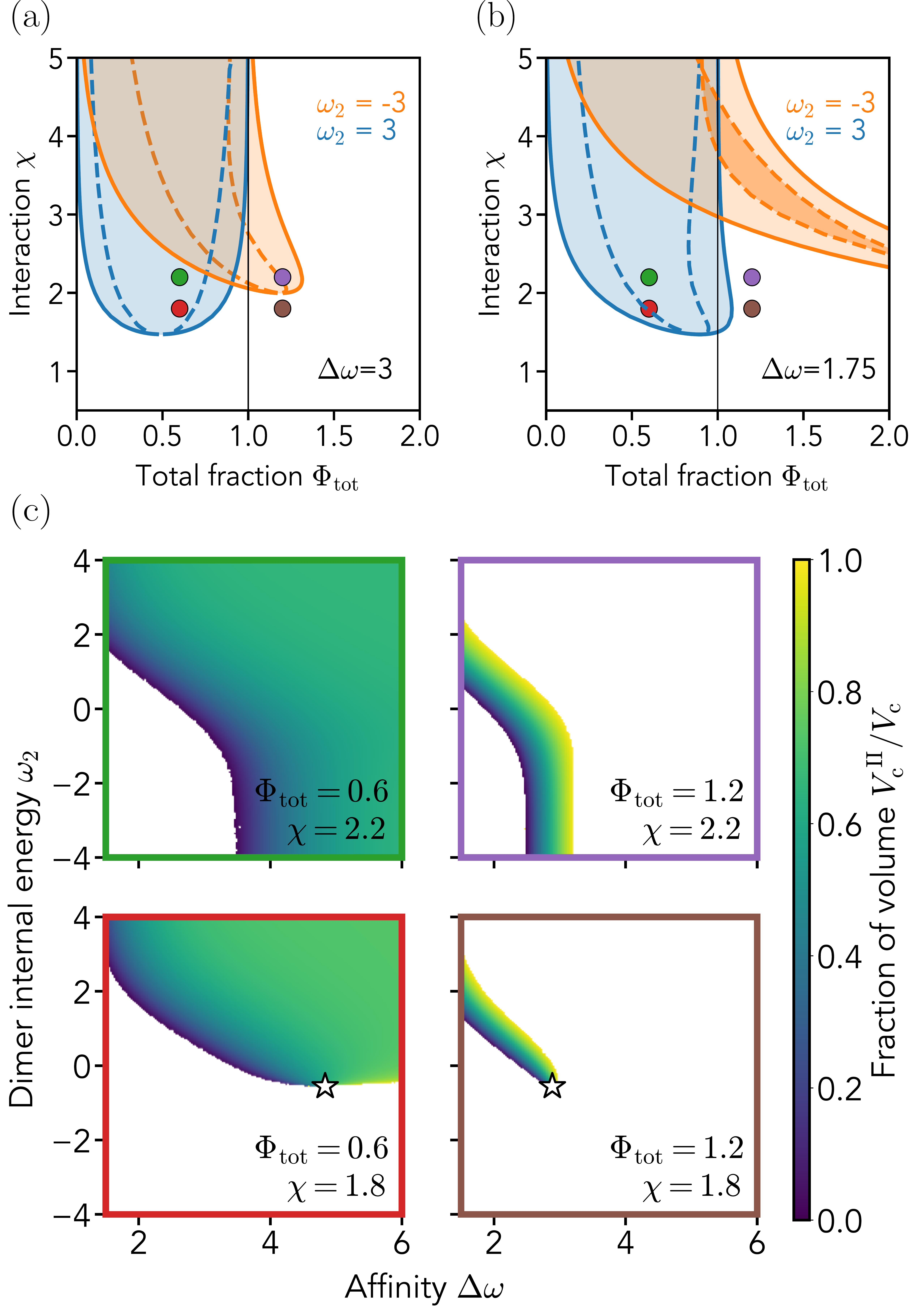}}

  \caption{\textbf{Dimerization and bulk sequestration jointly regulate compartment phase separation}
  (a)--(b) Phase diagrams as a function of the total fraction~$\PhiTot$ and interaction strength~$\chi$ for two dimerization energies $\omega_2$ and different compartment affinities $\Domega$.
The compartment can phase separate within the colored regions.
Dashed spinodal lines mark the stability of homogeneous states; the dark orange region in (b) indicates that both homogeneous states are stable.
 (c) Volume~$\Vc^\mathrm{II}$ of the dense phase normalized to compartment volume~$\Vc$ as a function of $\Domega$ and $\omega_2$ for various $\PhiTot$ and $\chi$ as indicated. %
Critical points are marked by stars.
 (a)--(c) Phase diagrams have been determined numerically; The size ratio is~$\eta=100$.
}
  \label{fig:phase_diags_fullmodel}
\end{figure}

We finally combine all effects to understand how phase separation in the compartment could be regulated. %
\figref{fig:phase_diags_fullmodel}a--b show examples for phase diagrams for various dimerization energies~$\omega_2$ and binding affinities~$\Domega$.
In essence, we find a combination of the effects obtained from the individual considerations above:
Favoring dimers (larger $\omega_2$) reduces the interaction~$\chi$ required for phase separation, similar to the results shown in \figref{fig:dimers+ph_sep}.
In contrast, bulk sequestration moves the binodal region to larger total fractions~$\PhiTot$, analogous to \figref{fig:phase_diags_phsep}.
This again implies a parameter regime where three states are stable (dark orange region in \figref{fig:phase_diags_fullmodel}b) for low $\Domega$.
The same regime also emerges when the dimerization energy~$\omega_2$ is reduced (see \figref{fig:dimer_binding}d) or $\eta$ is increased.
This suggests that the three parameters $\Domega$, $\omega_2$, and $\eta$ effectively control the tendency of the material to be in the compartment.

We find that all parameters in the system can be used to control whether the compartment phase separates and what fraction~$\Vc^\mathrm{II}/\Vc$ is occupied by the dense phase; see \figref{fig:phase_diags_fullmodel}c.
The fraction of volume $\Vc^\mathrm{II}/\Vc$ generally increases with increasing $\Domega$, $\omega_2$, $\PhiTot$, and $\chi$.
The plots also reveal various phase transitions, including re-entrant transitions where increasing a parameter (for instance $\Domega$ in the upper right panel) induces a transition from a dilute homogeneous phase to a phase separated state, and then a dense homogeneous state.

\floatsetup[figure]{style=plain,subcapbesideposition=top}

\section{Discussion}

Our equilibrium theory provides a comprehensive summary of the interplay of dimerization, compartment binding, and phase separation.
In the simplest cases, we can understand phase separation in the compartment by treating the bulk as a perturbation:
The bulk sequesters a certain fraction of material, lowering the available fraction in the compartment, which is relevant for phase separation.
The sequestration depends on the size ratio~$\eta$, the binding affinity~$\Domega$, the dimerization energy~$\omega_2$, and the interaction parameter~$\chi$.
However, since $\omega_2$ and $\chi$ also directly affect phase separation, there can be an intricate interplay between sequestration and phase separation, which for instance leads to the three stable states shown in \figref{fig:phase_diags_phsep}d and \figref{fig:phase_diags_fullmodel}b.
This multistability could in principle be observed in small compartments (e.g., lipid droplets~\cite{Kamatar2024}, the synaptomenal complex during meiosis~\cite{Durand2022}, or yeast vacuoles~\cite{Leveille2022}).

We showed that dimerization plays a double role since it lowers the interaction energy $\chi$ required for phase separation (see \figref{fig:dimers+ph_sep}a) and it increases the effective affinity toward the compartment (see \figref{fig:dimer_binding}).
Both these roles can either be interpreted as halving the entropy of molecules that are bound in dimers or as doubling enthalpic contributions when comparing monomers to dimers.
These effects amplify existing energy differences between subsystems, so that even small affinities can concentrate most of the material in a compartment (see \figref{fig:dimer_binding}d).
If multiple subsystems compete with each other, this mechanism allows to target the most favored subsystem reliably.
For these reasons, and because dimerization is ubiquitous, it is plausible that cells exploit dimerization as a regulation mechanism~\cite{Bland2023}.
Moreover, the same principles apply for higher-order oligomers, which are also known to affect protein localization in cells~\cite{Illukkumbura2023,Tschirpke2023_Liedewij,Rana2024}

The multistability of homogeneous states and the amplification by oligomerization vanishes when one subsystem is much larger than the other.
When the bulk is small, the compartment can be described using a canonical ensemble; the opposing limit corresponds to a grandcanonical ensemble, where the large bulk plays the role of a chemostat.
Our model reveals complex behavior between these classical limiting cases.
Similar quantitative dependencies on the size ratio of subsystems were found experimentally studying the PAR protein system~\cite {Hubatsch2019, Gessele2020} and the Min protein oscillations~\cite {Brauns2021_Nature,Brauns2022}, suggesting that it might be another important control parameter in cells.

Our equilibrium theory captures basic ingredients that allow regulating phase separation in a compartment.
The theory can be naturally extended to include multiple components, wetting phenomena~\cite{Zhao2021, Zhao2024}, mechanical degrees of freedom (like membrane fluctuations~\cite{Fournier2021}, membrane curvature~\cite{Tozzi2019,Yuan2021}, receptor binding~\cite{Lin2024}, or bulk elasticity~\cite{Qiang2024}), and activity (using open systems~\cite{Zamparo2021, Floris2022}, or driven reactions~\cite{Caballero2023, Kirschbaum2021, Zwicker2022}).
Those ingredients can help to regulate phase separation in the compartment, which could for instance explain polarity patterns in membranes~\cite {Burkart2022, Gross2019, Brauns2023_Liedewij}. %
Finally, our equilibrium theory will be the basis for thermodynamically-consistent kinetic theories that could describe the dynamic behavior of condensates in cells.

\begin{acknowledgments}
We thank Liedewij Laan, Nynke Hettema, Marieke Glazenburg , Zhiheng Wu, Nathan Goehring, Tom Bland, Marcel Ernst, and Kueyoung Kim for helpful discussions.
We gratefully acknowledge funding from the Max Planck Society and the European Union (ERC, EmulSim, 101044662).
GW acknowledges funding through a fellowship of the IMPRS for Physics of Biological and Complex Systems.
\end{acknowledgments}

\appendix

\section{Boundary coupling between compartment and bulk}
\label{sec:boundary_coupling}
A direct interaction between bulk and compartment through a boundary coupling can potentially affect the behavior of both subsystems.
If both subsystems are three-dimensional, the coupling is a surface contribution, which is negligible compared to bulk terms.
However, if the compartment is only two-dimensional it couples to the bulk everywhere and can have profound effects~\cite{Zhao2021}.

For simplicity, we only consider monomer fractions $\phi$ and $\psi$, and introduce the boundary coupling by expressing the overall free energy $F$ given by \Eqref{eqn:free_energy} as an integral over the two-dimensional compartment~\cite{deGennes1985, Ziethen2024, Cahn1977},
\begin{align}
	\label{eqn:bc_free_energy_total}
	\tilde F[\phi,\psi] = \int_{\Vc} \Bigl[ f_\tm{c}(\phi)  + f_\tm{bc}(\phi,\psi) \Bigr] \diff^2 r 
	\;.
\end{align}
The first term captures the internal energy density of the compartment, whereas the second term captures both the interaction and the internal energy density of the bulk, %
\begin{equation} \label{eq:bc_free_energy}
	f_\mathrm{bc}%
	=g\bigl(\phi, \left.\psi\right|_{z=0}\bigr)  + \int_0^{L_\mathrm{b}}\!\! \left[f_\mathrm{b}(\psi) + \frac{\kappa}{2} |\partial_z\psi|^2 \right]\mathrm{d}z
	\;,
\end{equation}
where $z$ is a coordinate perpendicular to the boundary reaching into the bulk of length~$L_\mathrm{b} = \Vb/\Vc$.
The coupling is described by a contact potential $g$, which depends on~$\phi$ and the bulk fraction~$\psi$ at the boundary at $z=0$.
Beside the bulk energy density $f_\mathrm{b}$, we also include a gradient term proportional to $\kappa$ to describe deformations of $\psi$ at the boundary.
The respective equilibrium profile follows from minimizing \Eqref{eq:bc_free_energy}, implying~\cite{Ziethen2024}
\begin{subequations} 
\begin{align}
	f_\mathrm{b}'(\psi) - \kappa \partial_z^2\psi & = f_\mathrm{b}'(\psi_\infty) \label{eq:min_conditions1} 
\\
	\kappa\left.\partial_z\psi\right|_{z=0} & = 
	\partial_\psi g\bigl(\phi, \left.\psi\right|_{z=0}\bigr)
	\;,  \label{eq:min_conditions2}
\end{align}
\end{subequations}
where $\psi_\infty$ is the bulk  fraction far from the boundary. %

The coupling can in principle induce wetting, pre-wetting, or a correction localized to the boundary, depending on the interactions in the bulk~\cite{deGennes1985}.
Since we here focus on systems that phase separate in the compartment, we exclude full and partial wetting in the bulk.
Yet, prewetting is in principle still possible, which could induce feedback between dense phases in the compartment and prewetted regions in the bulk~\cite{Zhao2021, Zhao2024}.
Prewetting typically requires fine-tuned parameters, so we do not investigate it in detail.
Instead, we here focus on the small changes in the bulk profile~$\psi$ induced by the coupling.
We thus expand $f_\mathrm{b}$ around $\psi_\infty$, so \Eqref{eq:min_conditions1} yields
$\psi(z) = \psi_\infty  - \alpha g_1(\phi) e^{-z/w}$, with a microscopic length $w = [\kappa/ f_\mathrm{b}''(\psi_\infty)]^{1/2}$.
The non-dimensional constant $\alpha=w k_\mathrm{B} T/(\kappa \nuc)$ follows from \Eqref{eq:min_conditions2}  with %
\begin{equation} \label{eq:contact_pot}
	 g\bigl(\phi, \left.\psi\right|_{z=0}\bigr) \approx \frac{k_\mathrm{B}T}{\nuc} \Bigl[g_0(\phi) + g_1(\phi) \left.\psi\right|_{z=0} \Bigr]
	 \;,
 \end{equation}
 where $g_0$ and $g_1$ are  series coefficients from an expansion in $\psi$.
 The system's state is thus quantified by $\phi(x, y)$ and $\psi_\infty$, similar to the model in the main text.
However, the coupling now induces a change  
 $\delta f_\mathrm{bc} =  f_\mathrm{bc}
 -L_\mathrm{b} f_\mathrm{b}(\psi_\infty)$ in the energy density, which we express as
\begin{equation}
	\delta f_\mathrm{bc} \approx \frac{k_\mathrm{B} T}{\nuc} \bigg[
			\bigg(\psi_\infty -\frac{f_\mathrm{b}'(\psi_\infty)}{f_\mathrm{b}''(\psi_\infty)}\bigg) g_1(\phi) -
			 \frac{\alpha}{2}g_1^2(\phi) %
	\bigg]
	\;,
\end{equation}
where we expanded  $f_\mathrm{b}$ for $\psi\approx\psi_\infty$, inserted the equilibrium profile~$\psi(z)$, and used $L_\mathrm{b} \gg w$. %
Assuming a dilute bulk, $f_\mathrm{b} \approx \frac{k_\mathrm{B}T}{\nub} [ \psi\ln\psi + \psi( \omega_\mathrm{b} -1 +\chi)]$, this becomes
\begin{equation}
	\delta f_\mathrm{bc} \approx
		-\frac{\nub}{\nuc} \bigl[f_\mathrm{b}(\psi_\infty) - f_\mathrm{b}(0)\bigr] g_1(\phi)
		- \frac{\alpha \kBT}{2\nuc}g_1^2(\phi)
	\;.
\end{equation}
After  integration over the boundary, see \Eqref{eqn:bc_free_energy_total}, the first term leads to a correction of the bulk free energy,
\begin{equation}
	\tilde F_\mathrm{b} = V_\mathrm{b}f_\mathrm{b} \bigg( 1 - \frac{g_1(\phi)}{\eta}\bigg)
	\;,
\end{equation}
so the coupling does not affect the bulk significantly when $g_1 \ll \eta$ (since distortions are localized to the boundary).

The coupling can have a stronger influence on the compartment.
To quantify this, we investigate the chemical potential $\tilde\mu_\mathrm{c}  = \nuc \delta \tilde F/\delta \phi$, which reads
\begin{align}
	\tilde\mu_\mathrm{c} = \muc 
			-\nub \bigl[f_\mathrm{b}(\psi_\infty)-f_\mathrm{b}(0)\bigr]g_1'(\phi)
		- \alpha \kBT g_1(\phi)g_1'(\phi) \;.
\end{align}
Comparing to \Eqref{eq:chempot_comp_phcomp}, this implies a  rescaling of $\Domega$ and $\chi$ in the simple case where  $g_1(\phi) = g_{1,0} +g_{1,1} \phi$,
\begin{subequations}
\label{eqn:coupling_rescaling}
\begin{align}
	\Domega'&\approx \Domega + \alpha  g_{1,1}\left(\frac{g_{1,1}}{2} + g_{1,0}\right)
\\
	\chi'&=\chi + \frac{ \alpha }{2} g_{1,1}^2
	\label{eqn:coupling_rescaled_chi}
	\;,
\end{align}
\end{subequations}
where we neglected a correction to $\Domega$ on the order of~$\psi$.
Consequently, a composition-dependent contact potential ($g_{1,1}\neq0$) amplifies phase separation in the compartment and promotes binding.

To understand the impact of the coupling onto the compartment more generally, we next study the spinodal condition $0=\tilde\mu_\mathrm{c}'(\phi)$, which reads
\begin{equation}
	\label{eqn:coupling_spinodal}
	\mu'_\mathrm{c}(\phi)  = 
			\bigl[\nub f_\mathrm{b}(\psi_\infty) + \alpha\kBT g_1(\phi) \bigr]g_1''(\phi) 
		+ \alpha\kBT g_1'(\phi)^2 \;.
\end{equation}
The right hand side collects all corrections, which could in principle affect the spinodal.
However, the first term is negligible compared to the interaction term resulting from \Eqref{eq:chempot_comp_phcomp} when $g_1''(\phi) \ll \chi / [\alpha g_1(\phi) + \nub f_\mathrm{b}(\psi_\infty)/ \kBT]$.
Here, the second term in the square bracket is always small compared to $1$, whereas the first term is small if $g_1(\phi) \ll \psi_\infty^{-1/2}$ since $\alpha \sim \psi_\infty^{1/2}$, so the whole correction is typically negligible if $g_1''(\phi) \ll \chi$.
The second term in \Eqref{eqn:coupling_spinodal}, related to the rescaling of $\chi$ given by \Eqref{eqn:coupling_rescaled_chi}, is negligible if $g_1'(\phi)\ll(\chi/\alpha)^{1/2}$.

Finally, the coupling can also affect the binding equilibrium between compartment and bulk, described by $\tilde{\mu}_\mathrm{c} = \mu_\mathrm{b}$ (since the correction to the bulk chemical potential is negligible).
The first correction given in \Eqref{eqn:coupling_spinodal} can be neglected since $f_\mathrm{b}g_1'(\phi) \ll f_\mathrm{b}'(\psi_\infty)$, and the second term is negligible when $g_1'(\phi)g_1(\phi) \ll \chi/\alpha$.

In summary, this shows that the coupling typically just rescales parameters and thus does not affect the qualitative results presented in the main text.
This rescaling is negligible if the contact potential $g_1(\phi)$ and its derivatives are small compared to $\alpha^{-1/2}\sim \psi_\infty^{-1/4} < \eta^{1/4}$, allowing for couplings of several $k_\mathrm{B} T$ for typical values ($\eta \approx 100$).

\section{Dimerization equilibrium} \label{app:dim_eq}
The free energy density of a single subsystem with a monomer fraction $\phi_1$ and dimer fraction $\phi_2$ reads
    \begin{multline}\label{eq:free_energu_dim_eq}
         \frac{f(\phi_1,\phi_2) \nu}{\kBT}   = 
         	\phi_{1} \ln(\phi_{1}) +\frac{\phi_{2}}{2} \ln(\phi_{2})   + (1-\phi_{\tm{}}) \ln(1-\phi_{\tm{}})\\ +\omega_{1}\phi_{1} + \omega_{2} \phi_{2}  \;.
    \end{multline}
Combining particle conservation, $\phi_1+\phi_2=\phi$, with the dimerization equilibrium, $\partial f/\partial \phi_1 = \partial f/\partial \phi_2$,~\cite {Canalejo2019}
\begin{align}\label{eq:dim_eq}
\phi_1 & = \frac{-1 + \sqrt{1+4 K_\tm{d} \phi}}{2 K_\tm{d}} 
&
\phi_2 &= \phi - \phi_1
\end{align}
with the dimerization constant $K_\tm{d}=\phi_2/\phi_1^{2}$ given by $K_\mathrm{d}=e^{2(\omega_{2}+1)}$.
Inserting these relations in \Eqref {eq:free_energu_dim_eq} results in an effective free energy that only depends on $\phi$, implying the effective chemical potential $\mu=\nu\partial f/\partial \phi$. 
The spinodal condition, $f''(\phi)=0$, then implies
\begin{equation}
    \chi_{\tm{sp}}=\frac{1}{4\bar\phi} \bigg(\frac{1}{ \sqrt{1+4 K_\tm{d} \bar\phi}}+\frac{1+\bar\phi}{1-\bar\phi} \bigg)
    \;.
\end{equation}

\section{Limiting cases of dimerization equilibrium in coupled subsystems}
\label{app:transition_regimes}
We first consider the limiting case of only monomers, focusing on dilute systems for simplicity.
In equilibrium, the solute compartment fraction~$\bar\phi$ then reads
\begin{equation}
\bar{\phi}_{\mathrm{mono}}=\frac{\Phi_{\mathrm{tot}}e^{\Domega}}{\eta+e^{\Domega}} \;.
\end{equation}
In the case of dilute systems filled with dimers, we find
\begin{equation}
\bar{\phi}_{\mathrm{dim}}=\frac{\Phi_{\mathrm{tot}}e^{2\Domega}}{\eta+e^{2\Domega}} \;.
\end{equation}
In the intermediate regime, we consider the compartment to only contain dimers, whereas the bulk only has monomers, leading to
\begin{multline}
\bar{\phi}_\mathrm{int} = \frac{e^{-(1+2(\omega_2+\Domega)}}{2} \bigg[\eta^2+2e^{1+2(\omega_2+\Domega}\PhiTot \\ - \eta \sqrt{\eta^2+4e^{1+2(\omega_2+\Domega} \PhiTot}] \bigg] \:.
\end{multline}
The transitions between these limiting cases take place when $\bar{\phi}_{\mathrm{mono}}=\bar{\phi}_\mathrm{int}$ and $\bar{\phi}_\mathrm{int}=\bar{\phi}_{\mathrm{dim}}$, leading to the two respective transition points
\begin{align}
 \Phi_{\mathrm{tot,mono}} &= \frac{e^{\Domega}+\eta}{e^{1+2\omega_2+\Domega}}
&
 \Phi_{\mathrm{tot,dim}} &= \frac{e^{2\Domega}+\eta}{e^{1+2 \omega_2}}  \:.
 \end{align}

\section{Phase separation in a compartment exchanging material with the bulk} \label{app:ph_sep_comp}

The free energy of a phase separated compartment in contact with a homogeneous bulk reads
 \begin{equation} \label{eq:free_energy_total}
     F= V_\tm{c}^{\tm{I}}f_\tm{c}(\phi^{\tm{I}})+V_\tm{c}^{\tm{II}}f_\tm{c}(\phi^{\tm{II}})+V_\tm{b} f_\tm{b}(\psi) 
     \;,
 \end{equation}
 subject to the conservation laws
$V_\tm{c}^{\tm{I}}+V_\tm{c}^{\tm{II}}=V_\tm{c}$,
$V_\tm{c}^{\tm{I}} \phi^{\tm{I}}+ V_\tm{c}^{\tm{II}} \phi^{\tm{II}} = V_\tm{c} \phi$, and
$\eta \psi + \phi = \Phi_{\text{tot}}$,
where $\eta =V_\tm{b} \nu_\tm{c}/V_\tm{c} \nu_\tm{b}$.
The necessary conditions for the minimum of $F$ thus read
    \begin{align}
    \label{eq:binodal_cond}
	\frac{V_\tm{b}}{V_\tm{c} \eta}f'_\tm{b}(\psi) 
     =  f'_\tm{c}(\phi^{\tm{I}})
       = f'_\tm{c}(\phi^{\tm{II}})
      =   \frac{f_\tm{c}(\phi^{\tm{I}})-f_\tm{c}(\phi^{\tm{II}})}{\phi^{\tm{I}}-\phi^{\tm{II}}}
	\;.
    \end{align}
We can simplify these conditions to
\begin{equation} \label{eq:binodal_cond_sym}
	\tilde{f}_\tm{b}'(\psi)=\tilde{f}_\tm{c}'(\phi^{\tm{I}})=\tilde{f}_c'(\phi^{\tm{II}})=0
\end{equation}
by defining effective free energy densities,
\begin{subequations} \label{eq:simp_free_en}
\begin{align}
    \tilde{f}_\tm{c}& = \frac{\kBT}{\nu_\tm{c}} \Bigl(\phi \ln(\phi) +(1-\phi) \ln(1-\phi)  + \chi \phi(1-\phi)\Bigr) 
    \\
    \tilde{f}_\tm{b} & = \frac{\kBT}{\nu_\tm{b}} \Bigl(\psi\ln(\psi) + (\Domega +\chi-1)\psi \Bigr)  \label{eq:simp_free_en_b} \;,
\end{align}
\end{subequations}
with $\Domega = \omega_\tm{b} - \omega_\tm{c} $.
The first identity in \Eqref {eq:binodal_cond_sym} implies
\begin{equation} \label{eq:volfrac_cond}
  \frac{V_\tm{c}^{\tm{I}}}{V_\tm{c}}= \frac{1}{\phi^{\tm{I}}-\phi^{\tm{II}}}\Bigl( \Phi_{\text{tot}} -\phi^{\tm{II}} - \eta e^{-\Domega -\chi} \Bigr)  \;,
\end{equation}
and we can determine $\phi^{\tm{I}}$ and $\phi^{\tm{II}}$ from the two solutions of the second identity in \Eqref {eq:binodal_cond_sym},
\begin{equation}
	\chi_\mathrm{bin}= \frac{1}{2\phi-1}\ln\left(\frac{\phi}{1-\phi}\right)
	\;.
\end{equation}
Consequently, the coexisting fractions only depend on $\chi$, implying they can be read off the canonical phase diagram. %
The region of $\PhiTot$ where phase separation happens is then given by $V_\tm{c}^{\tm{I}} \in (0, \Vc)$, so that \Eqref{eq:volfrac_cond} implies
\begin{subequations}
\begin{align}
    \PhiTot^{\mathrm{min}} & =\phi^{\tm{I}} + \eta e^{-\Domega -\chi} \\
    \PhiTot^{\mathrm{max}} & =\phi^{\tm{II}} + \eta e^{-\Domega -\chi} \;. 
\end{align}
\end{subequations}
Since the chemical potentials are all constant, the bulk fraction~$\psi= \exp(-\Domega - \chi)$ is a constant  in the region where the compartment phase separates.

The compartment is homogeneous in equilibrium when $\mu_\tm{c}(\phi^*)=\mu_\tm{b}(\psi^*)$.
 To see whether this state is stable, we test whether the perturbed state $\phi(\vect r)=\phi^*+\delta\phi(\vect r)$ increases the total free energy
\begin{equation}
\label{eq:total_free_energy}
    F[\phi,\psi] = V_\tm{b}f_\tm{b}(\psi) + \int_{V_\tm{c}}  f_\mathrm{c}(\phi) \tm{d}V_\tm{c}  \;,
\end{equation}
assuming a homogeneous bulk.
The first variation of the free energy, $\delta F[\phi] = F[\phi] - F[\phi^*]$, reads
\begin{multline} \label{eq:free_en_expanded}
\delta F[\phi]  =\int_{V_\tm{c}} \bigg[ f_\tm{c}'(\phi^*)\delta \phi + \frac{f''_\tm{c}(\phi^*)}{2} \delta \phi^2\bigg] \tm{d}V_\tm{c} \\ + V_\tm{b} \left( f_\tm{b}'(\psi^*) \delta \psi +\frac{f''_\tm{b}(\psi^*)}{2}\delta \psi^2 \right)  \;.
\end{multline}
Material conservation, given by \Eqref{eq:conservation}, implies
\begin{equation}
    \delta \psi(t) = - \frac{1}{\eta} \int_{V_\tm{c}} \frac{\delta \phi(r,t)}{V_\tm{c}} \tm{d}V_\tm{c}  \;.
\end{equation}
The terms proportional to $\delta\phi$ and $\delta\psi$ thus cancel in  \Eqref{eq:free_en_expanded}.
We express remaining terms in Fourier space,
\begin{equation}
    \delta F[\hat{\phi}]  = V_\tm{c} \int \diff^n q \left[\frac{f''_\tm{c}(\phi^*)}{2} \hat{\delta\phi}(\vect q)^2 \right] + \frac{V_\tm{b}}{2 \eta^2}  f''_\tm{b}(\psi^*) \hat{\delta \phi}(\vect 0)^2  \;. 
\end{equation}
If the bulk is stable, $f''_\tm{b}(\psi^*)>0$, the second term increases $\delta F$, implying that the homogeneous state is stable if the first term is positive, i.e., if $f''_\tm{c}(\phi^*)>0$.

To estimate when the bulk could phase separate, we check whether its fraction~$\psi$ lies within the binodal region.
The gray region in Fig.~\ref{fig:phase_diags_phsep} marks such regions.

\bibliography{compartment_draft}%

\end{document}